\documentstyle[12pt,epsf]{article}
\newcommand{\be}{\begin{equation}}
\newcommand{\ee}{\end{equation}}

\begin{document} YITP-00-73 
\hspace{10cm}
\today
\\
\vspace{3cm}
\begin{center} {\LARGE   New results for the quantum supersymmetric kink } 
\\  \vspace{2cm} {Alfred Scharff Goldhaber\footnote{e-mail:
goldhab@insti.physics.sunysb.edu},  Andrei Litvintsev\footnote{e-mail:
litvint@insti.physics.sunysb.edu}  and Peter van
Nieuwenhuizen\footnote{e-mail: 
vannieu@insti.physics.sunysb.edu}\\
 { \it C.N.Yang Institute for Theoretical Physics, SUNY at Stony Brook},
\\ {\it Stony Brook, NY 11794 } }
\abstract{   
We review our work on computations of the quantum corrections to the mass and the central charge
of the susy kink. For the mass corrections, we 
find that the widely used momentum cut-off scheme gives an incorrect result, but we deduce through
smoothing of the cut-off an extra term in the
mass formula, which produces the correct result.
We discover the importance of boundary effects for the mode number cut-off regularization
scheme. We introduce the notion of delocalized boundary energy. We discuss two discrete 
$Z_2$ symmetries and their importance to the mode number approach. For the
central charge corrections, we use momentum cut-off regularization with two 
cut-offs, one for propagators and another for Dirac delta functions. We then compute the quantum
anomaly in the central charge, and find that it restores the BPS bound at the one-loop 
level if the two cut-offs are equal.  
 }
\end{center}
\vspace{4cm}
Talk at the E. S. Fradkin Memorial Conference, June 2000 
\newpage

The problem of quantum corrections to the mass of the bosonic and susy kink was intensively studied
in many papers in the 1970's and 1980's (see \cite{raj} and references therein). 
In 1997, with the advent of extended particles in string theory, the issue of quantum corrections to 
extended objects, and the issue of the BPS bound at the quantum level, was brought back into focus in ref. \cite{rebhan}.  
This paper emphasized the contradictory results 
coming from the two major regularization approaches, namely the momentum and mode number cut-off schemes.
Another issue emphasized in \cite{rebhan} was that the BPS bound did not seem saturated 
at the quantum level, though clearly it holds classically. 
Given the discrepancies between the results of these
two schemes, in \cite{misha} another regularization scheme for the evaluation of the quantum mass of a soliton was developed,
which we have called derivative regularization.
According to this method one first evaluates the sums $\frac{\partial}{\partial m} \sum \omega_n$ 
which are better convergent than $\sum \omega_n$. 
In \cite{misha} it was also proposed that there is an anomaly which is responsible for the
BPS bound saturation on the quantum level. 
The results of \cite{misha} on the quantum mass of a (susy) kink were confirmed by the phase shift approach of \cite{graham}
who also presented a calculation of the central charge.
The anomaly was discovered later by \cite{shifman} and with this anomaly the saturation of the BPS bound was explained.
These new results did not answer the question
``what was wrong with the old schemes ?''  Recently we reanalyzed in \cite{we,me,fred} the old schemes. In this
contribution we present the results of this analysis  and
show that these schemes require some modifications.
These modifications have a physical motivation behind them and lead to correct results.

We study the (susy) kink with the Lagrangian
\be {\cal L}= -\frac{1}{2} (\partial_\mu \phi)^2 - \frac{1}{2} U^2(\phi) -
\frac{1}{2} \bar{\psi} \gamma^\mu \partial_\mu \psi - \frac{1}{2} c
\frac{dU}{d \phi} \bar{\psi} \psi
\label{Lagrangian}
\ee  where  $U[\phi(x)] = \sqrt{\frac{\lambda}{2}} \left(
\phi^2 - 
\frac{\mu_0^2}{\lambda} \right) $ and $\bar{\psi}=\psi^{\dagger}  i
\gamma^0$.
The field equations allow the static solution $\phi(x) = \frac{\mu}{\sqrt{\lambda}} \tanh \frac{\mu x}{\sqrt{2}}$, $\psi=0$, the kink, see fig.1.
To discretize the spectrum of fluctuations we introduce boundaries. As we will describe later, there is 
usually extra energy created by the boundaries. One needs to subtract this energy in order to obtain the correct mass of the (susy) 
kink \cite{schonfeld,shifman}.
\\ ~ \\ $$ \epsfbox{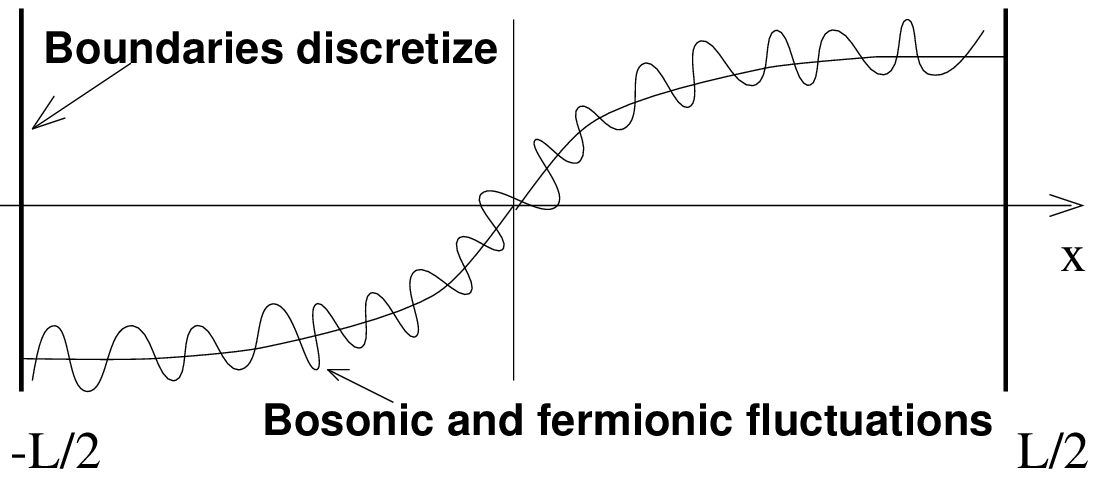} $$ Fig1. {\it The kink solution and quantum fluctuations around it.} \\ \\
The kink solution satisfies the BPS equation $\partial_x \phi + U =0$. One may define the central charge from the algebra
of the susy charges, and it is well known that the BPS bound is saturated at the classical level. 
To compute the quantum one-loop corrections we first renormalize the theory by replacing $\mu_0^2$ by $\mu^2+\delta \mu^2$, where 
the mass 
counter term $\delta \mu^2$ in the susy case is fixed by requiring that it cancels the tadpole
graphs 
\be 
\left( \delta \mu^2 \right)_{susy} = 
\frac{\lambda \hbar}{4 \pi} \int_{-\Lambda}^{\Lambda}
\frac{dk}{\sqrt{k^2+m^2}}; \hspace{0.5cm}
\epsfxsize 1.3in \raisebox{-0.4cm}{\epsfbox{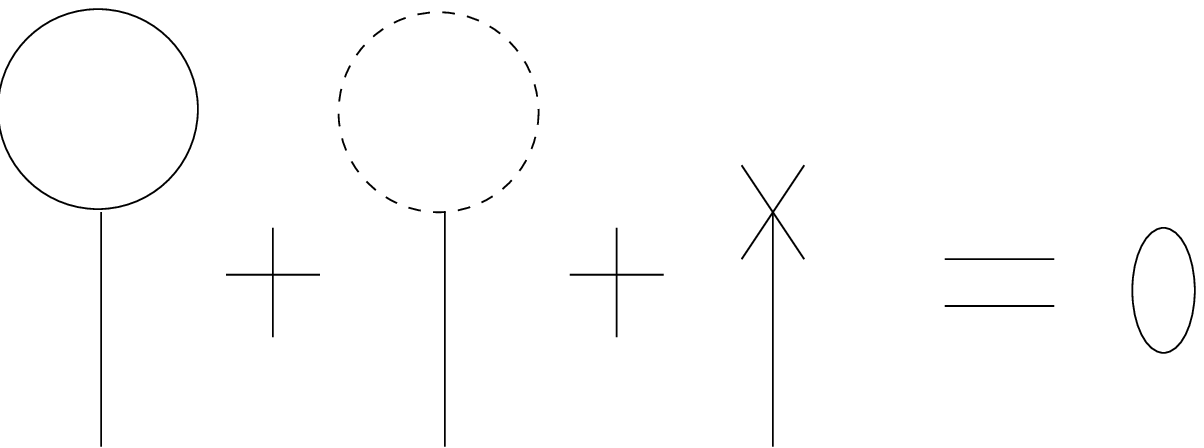}}
\ee
If one plugs the expression for the kink background into the renormalized 
Lagrangian, one obtains a counterterm of order $\hbar$, namely $ \Delta M =\frac{m}{\lambda} \left( \delta \mu^2 \right)_{susy}$.
Thus if we define the mass of the kink as
\be
M= \langle {\rm kink} | \hat{H} | {\rm kink} \rangle -  \langle {\rm trivial} | \hat{H} | {\rm trivial} \rangle 
\ee
then the expression for the one-loop correction to the mass of the susy kink is
\be
\label{summa}
M^{(1)}=\frac{\hbar}{2} \left( \sum \left[ \omega_n^b- \omega_n^{b,(0)} \right] \right) - \frac{\hbar}{2} \left( \sum \left[ \omega_n^f- 
\omega_n^{f,(0)} \right]  \right) + \Delta M
\ee

The computation of this expression is rather tricky. 
The first correct method to obtain the one loop mass correction is to begin with a kink-antikink
background, and to compute the spectrum of fluctuations around this background. 
By iterating the Dirac equation for $\psi = \left( \begin{array}{c} \psi_1 \\ \psi_2  \end{array} \right)$, all field 
equations become Schor\"odinger equations with the potentials shown in figure 2.
\\ \\ 
\epsfbox{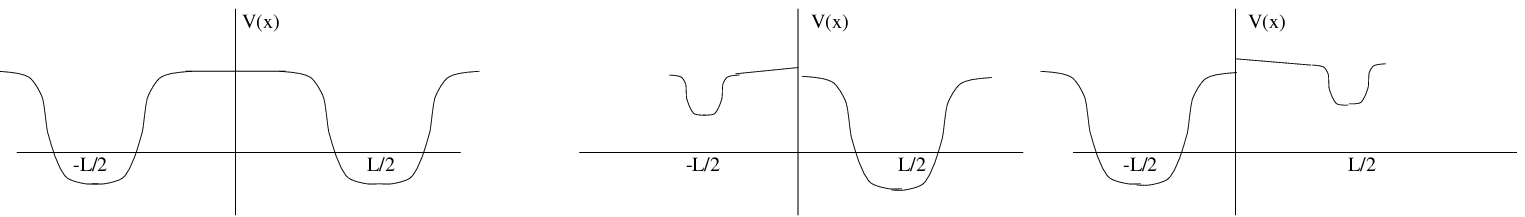} 
  
\hspace{1cm} $V(x)$  for $\eta$ \hspace{3cm} $V(x)$ for $\psi_1$ \hspace{3cm}
$V(x)$ for $\psi_2$ \\ \\ Figure 2. {\it Potentials for the bosonic and fermionic
fluctuations in  the kink-antikink system.} \\ 

Suppose
one starts with the constant solution $\phi_{cl}= \mu / \sqrt{\lambda}$, and starts pulling the field $\phi$
near zero down to $\phi_{cl}= - \mu / \sqrt{\lambda}$. One creates in this way a kink-antikink system
with the same boundary conditions in the kink-antikink sector as in the trivial sector,
and the mass of the kink is then
one-half the mass of the kink-antikink system \cite{schonfeld}.
The results are independent of the choice of boundary conditions.
A correct method for a single kink is the derivative regularization of
\cite{misha}. The idea is to take $\frac{\partial}{\partial m} M^{(1)}$ and then, after computing the sums, 
to fix all ambiguities by imposing the
renormalization condition $M^{(1)}(m=0)=0$ and to integrate to get $M^{(1)}=m \frac{\partial}{\partial m} M^{(1)}$. 
Again we must use the same boundary conditions in both sectors.
The answers obtained by this method are
\be
M^{(1)}(bos) = - m \hbar \left( \frac{3}{2 \pi} - \frac{\sqrt{3}}{12} \right) 
\hspace{2cm}
M^{(1)}(susy) = - \frac{m \hbar}{2 \pi}
\ee 
This approach gives the correct result whenever the boundary conditions lead
to a finite result for (\ref{summa}), but it does not explain what was wrong with the other approaches. 

The problems of the momentum cut-off were solved recently in \cite{we}. We introduced a smooth
function $f(k,\Lambda) \sim \Theta(\Lambda-k)$ into the quantization conditions:
\be
kL + f(k,\Lambda) \delta(k) = 2 \pi n
\ee
where the phase shifts $\delta(k)$ are due to the reflectionless kink potential\footnote{
The kink model is member of a whole class of reflectionless potentials which share many special properties. For a nice 
discussion see \cite{graham}. 
} for the
fluctuations. The purpose of $f(k, \Lambda)$ is to make a smooth transition
between the  nontrivial quantization conditions at small $k$ to the trivial 
conditions for $k> \Lambda$. 
This modification leads to new terms in the expression for the density
\be
\label{plotnost}
\frac{dn}{dk} = \frac{L}{2 \pi} + f(k,\Lambda) \delta^\prime (k) + f^\prime (k, \Lambda) \delta(k)
\ee
The term $f \delta^\prime(k)$ truncates the infinite sums in a natural way, and this kind of truncation 
is widely used in the Casimir effect. 
The last term is absent in the naive approach and it is this term which is responsible for 
obtaining the correct answer (which follows from using the spectral density (\ref{plotnost})
to transform the sums of (\ref{summa}) into integrals).

Once the traditional momentum cut-off for the mass corrections was rehabilitated, we turned \cite{we,me} to 
the direct computation of the central charge anomaly with momentum cut-off regularization.  
The classical Noether current for supersymmetry is 
\begin{equation}
j^\mu = -( \partial_\nu \phi) \gamma^\nu \gamma^\mu  \psi - U 
\gamma^\mu \psi
\end{equation}
By using equal-time canonical (anti) commutation relations we can compute $Z$ from 
\be
\{ Q^\alpha, \bar{Q}_\beta \} = 2 i (\gamma^\nu)^\alpha_{\hspace{1.5mm}\beta} 
P_\nu- 2i 
(\gamma^5)^\alpha_{\hspace{1.5mm} \beta} Z
\ee 
The expression for the 
central charge is  
\be 
Z_{bos}=\int_{-L/2}^{L/2} dx \int_{-L/2}^{L/2} dx^\prime \left\{ U[\phi(x)]
 \partial_{x^\prime} \phi(x^\prime) + U^\prime[\phi(x)] \bar{\psi}(x^\prime ) \gamma^1 \psi(x) \right\} \hat{\delta}(x-x^\prime)
\ee
Formally the first term is a total derivative and the second term vanishes because $\psi$ is a Majorana spinor,
but one should use regularization.
To regulate this expression we 
impose a momentum cut-off $\Lambda$ on the propagators, and a cut-off $K$ on the Dirac delta functions which follow from 
the equal-time canonical commutation relations,
where 
\be
\langle \eta(x) \eta(x^\prime) \rangle = \int_{-\Lambda}^\Lambda \frac{dq}{2 \pi} \frac{e^{iq(x-x^\prime)}}{\sqrt{q^2+m^2}}; \hspace{1cm}
\hat{\delta}(x) = \int_{-K}^K \frac{dq}{2\pi} \exp(iqx)
\label{dp1}
\ee
Expanding $\phi$ into $\phi_K+\eta$ where $\eta$ is the quantum field, the terms quadratic in $\eta$ are of the
form $\frac{1}{2} U^{\prime \prime} \langle \eta^2 (x) \rangle \partial_{x^\prime} \phi_K (x^\prime ) + U^\prime 
\langle \eta(x) \partial_{x^\prime} \eta(x^\prime) \rangle $. The former is canceled by the mass renormalization
whereas the latter is a typical anomaly effect: it is formally zero by symmetric integration if one uses the unregulated delta
function, but with the truncated delta function one produces a kind of point splitting and the contribution is nonzero and finite.
Shifman, Vainshtein and Voloshin \cite{shifman} have shown that there is indeed an anomaly in the current multiplet of which $Z$ 
is part. 

Taking the vacuum expectation value of the regulated central charge we can easily
obtain the quantum correction to $Z$.
We found that it only depends whether $\Lambda>K$, $\Lambda=K$ or $\Lambda<K$. 
Physically one needs $\Lambda=K$ because $(\Box +m^2) \hat{G}(x,x^\prime) = \hat{\delta}(x-x^\prime)$ where $\hat{G}(x,x^\prime) $
is the regulated propagator.
The propagator in the kink background can be written as a sum of the flat-space propagator, 
the propagator with one insertion of the background,
two insertions etc., see figure 3. 
\\ $$
\epsfbox{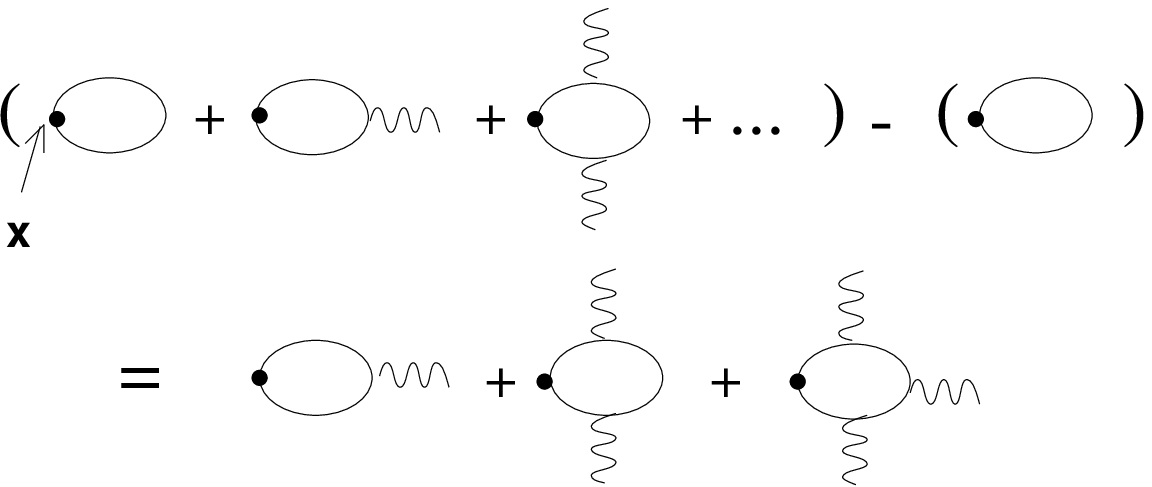}  $$
Figure 3. {\it The insertions in the loops are due to the kink background.} \\ \\
We found that only the flat space propagator gives a nonzero result, confirming that one is dealing with an anomaly
which is a local effect independent of the background.
For $\Lambda=K$ we found 
that the BPS 
bound is saturated at one loop \cite{we}. 
[In \cite{shifman} the local $Z$ density and energy density around the kink were studied.]

The BPS bound should be saturated at the quantum level if multiplet shortening occurs. Initially it was believed that
irreducible non-BPS multiplets for $N=(1,1)$ susy in $1+1$ dimensions (with $H \ne Z$ ) contained 4 states and BPS multiplets
(with $H=Z$) only 2 states. Then it was noted in \cite{we} that non-BPS multiplets contain only 2 states. Finally, \cite{shifman} and
later \cite{fred} showed that BPS multiplets contain only one state. This state is of the form $(1+c_0) | \Omega \rangle$ where
$c_0$ is the zero mode of the 2-component Majorana fermion in a kink background. It may seem strange to have a state which is half
bosonic and half fermionic, but note that in $1+1$ dimensions the distinction between bosons and fermions is less clear. Since
there is multiplet shortening, after all, one should find that $\langle H \rangle = M$ equals $\langle Z \rangle$ and that is now
indeed the case.
Though the BPS multiplet is 
really one dimensional, there exist two gauge copies of it,
connected by the discrete $Z_2$ gauge symmetry \cite{fred}. Our results are in accord with
the results of \cite{shifman,ritz}.

The two discrete $Z_2$ symmetries $\psi \to -\psi$, and $\phi \to -\phi$ with $\psi_1 \leftrightarrow \pm \psi_2$ 
play an important role in the computation
of the quantum corrections by the mode number regularization scheme.
We started with a kink-antikink system (see figure 4), following \cite{schonfeld}.
\\ $$
\epsfbox{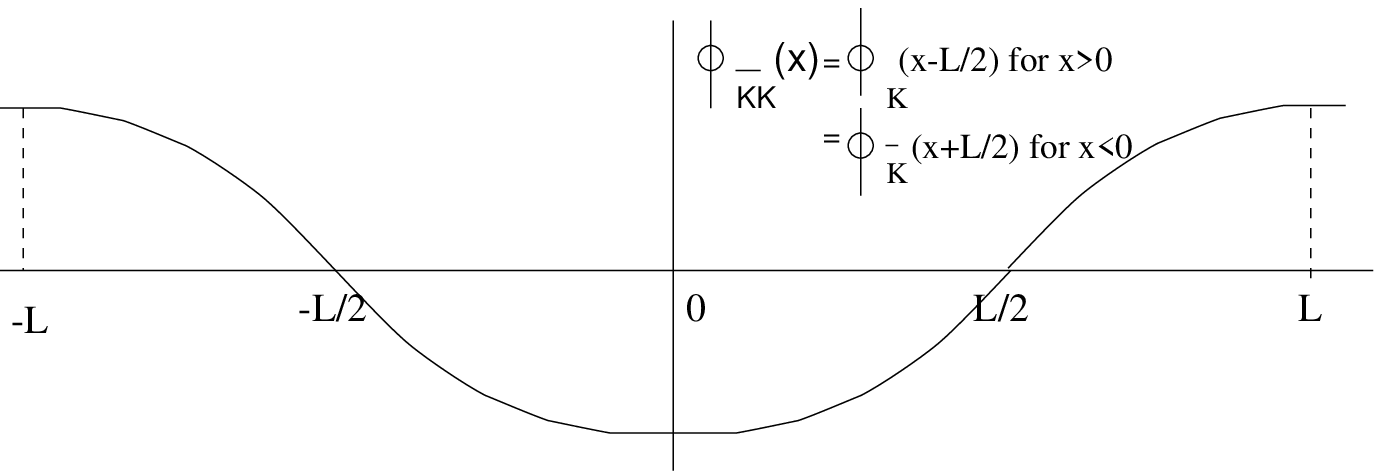} $$
Figure 4. {\it The kink-antikink configuration; there is a cusp at $x=0$.} \\ \\
This system  is topologically in the trivial sector (we have also
developed in \cite{fred} a susy version of 
Manton and Samols' sphalerons-on-a-circle model, see \cite{manton}, which lead to the same conclusions). 
One may use any (for example periodic) 
boundary conditions for the kink-antikink system and then deduce the set of boundary conditions needed for a single kink 
by looking at what happens in the middle. 
\\ $$
\epsfbox{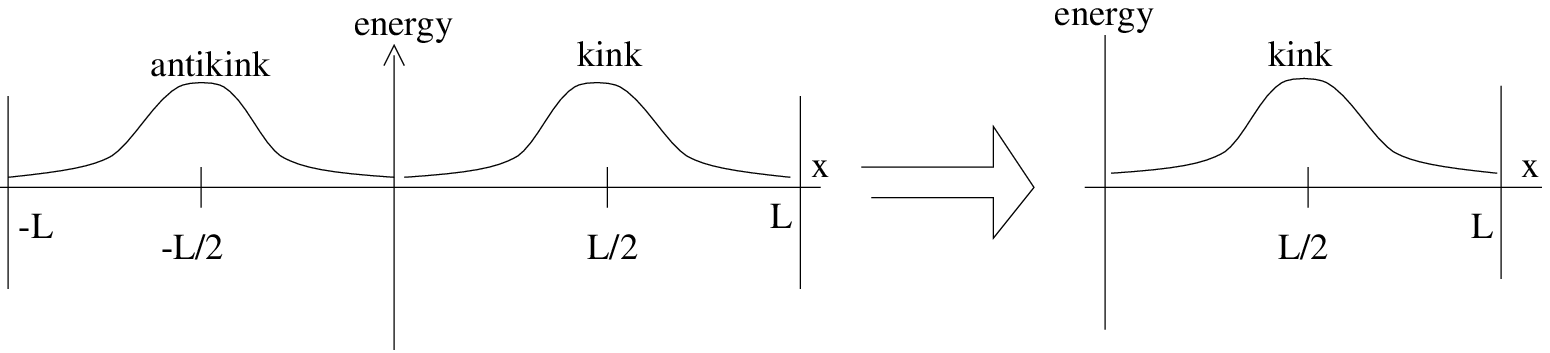} $$
Figure 5. {\it The boundary conditions for a single kink can be deduced from the boundary conditions of the kink-antikink system system.} \\ \\
Then,
if one averages over this set, one automatically obtains the correct answer. But why would it be necessary to average?
It turns out \cite{fred} that for a fixed set of boundary conditions for a single kink 
(namely, the same boundary conditions in the kink sector as in the trivial sector)
there is always a boundary energy, see figure 6.
\newpage
$$
\epsfbox{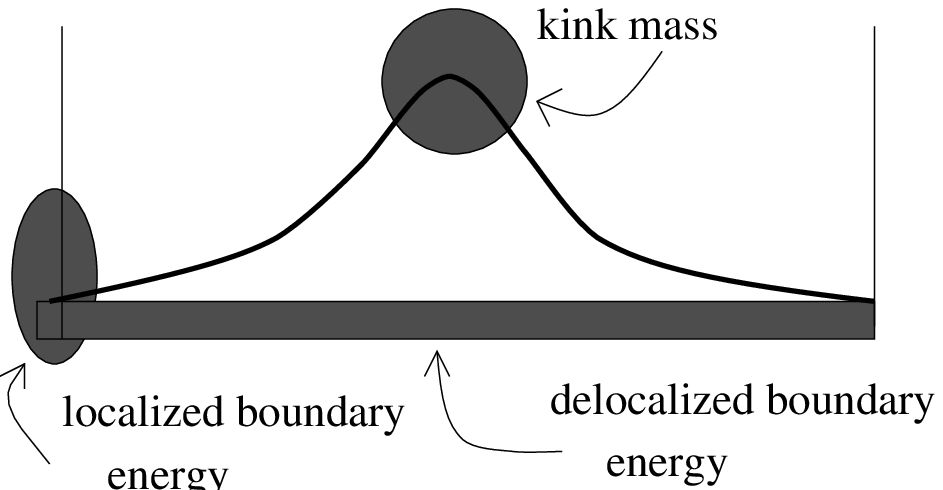}  $$
Figure 6. {\it The total energy density in the kink background.} \\ \\
For certain boundary conditions, one obtains standing waves and then
the boundary conditions are ``visible'' for the kink; then there is localized boundary energy in the kink sector.
For other boundary conditions, one finds plane waves as solutions, and then 
the boundary conditions are ``invisible'' for the kink but there can be delocalized boundary energy.
Boundary conditions which are invisible in the kink sector are visible in the trivial sector and vice-versa.
For a fixed set of boundary conditions, making a twist (i.e. exchanging $\psi_1$ and $\psi_2$ and replacing
$\phi \to -\phi$) at one side of the kink produces a current which drives half of a fermionic degree of freedom 
away from the kink to the boundary, see figure 7. This loss of half a degree of freedom was found to occur for Dirac fermions
in \cite{gw} and explained the discovery of \cite{jackiw} that a kink carries one half unit of fermion charge.  
\\$$
\epsfbox{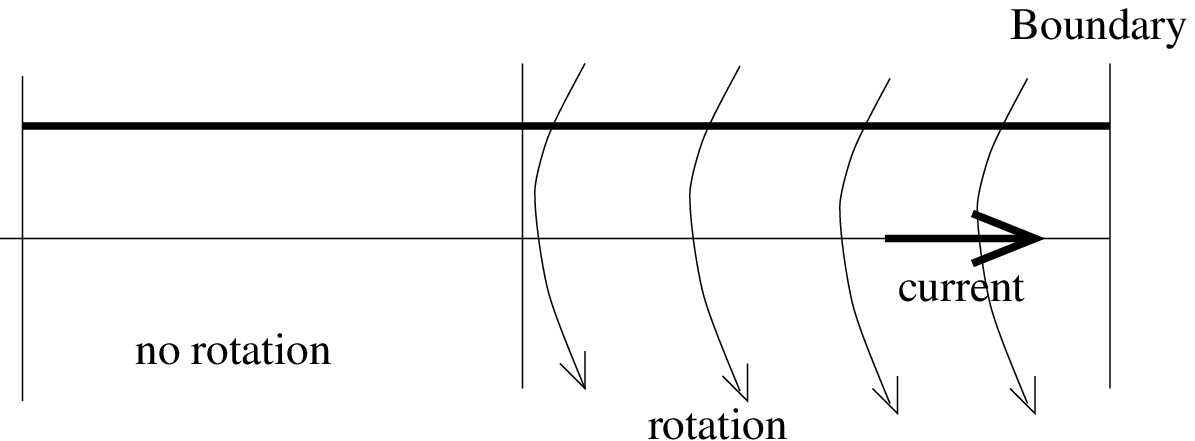}  $$
Figure 7. {\it The chiral rotation causes a current which leads to the accumulation of half the degree of freedom at the boundary.} \\ \\
On the other hand,
the $Z_2$ gauge symmetry guarantees that for a gauge invariant set of boundary
conditions, such as, for example, the set ``periodic+anti-periodic+twisted periodic+twisted anti-periodic,''
all boundary energies cancel. The chiral rotation of figure 7
produces from, say, the periodic conditions in one sector, both the twisted periodic and twisted anti-periodic
conditions in the other sector, which is due to the fact that one must complement
the chiral rotation by a gauge transformation to keep our fermions real. 
 
Thus we have not only rescued the oldest and most popular methods for computation of quantum corrections, 
mode number and momentum regularization, but we have also discovered the importance of boundary effects
for these computations. We have realized that there are two discrete $Z_2$ symmetries and that the failure to 
take a gauge invariant set of boundary conditions for the computation may even result in the appearance of
delocalized boundary energy, which would mean the failure of the cluster decomposition principle.

\end{document}